\documentclass[12pt]{article}
\usepackage[utf8]{inputenc}

\usepackage{color,xcolor, url, float,  verbatim, changepage}
\usepackage[linesnumbered,vlined,ruled]{algorithm2e}
\usepackage{setspace}
\usepackage{array,supertabular}
\usepackage{subcaption} 
\usepackage{mathtools,amsmath,amssymb} 
\usepackage{longtable} 
\usepackage{lmodern}
\usepackage[T1]{fontenc}
\usepackage{textcomp}
\usepackage{booktabs}
\usepackage[square,sort,comma,numbers]{natbib}

\usepackage[margin = 1in]{geometry}

\usepackage{amsthm}
\newtheorem{proposition}{Proposition}

\newtheorem{definition}{Definition}[section]

\def\halmos{\mbox{\quad$\square$}}

\DeclareMathOperator*{\argmin}{argmin}

\newcommand{\bt}[1]{\textcolor{black}{#1}}

\usepackage{xcolor}
\usepackage{soul}
\usepackage{authblk}

\title{Coordinating Resource Allocation during Product Transitions Using a Multifollower Bilevel Programming Model}
\author{Rahman Khorramfar\thanks{Corresponding author (khorram@mit.edu)\\
$\hspace{12cm} $ Rahman Khorramfar is affiliated with MIT Energy Initiative (MITEI) and Laboratory for Information $\&$ Decision Systems (LIDS), Cambridge, MA; Osman Ozaltin and Reha Uzsoy are affiliated with Edward P. Fitts Department of Industrial and Systems Engineering at NC State University, Raleigh, NC; and Karl Kempf is affiliated with Intel Corporation, Chandler, AZ \hspace{10cm} 
The authors acknowledge support from the National Science Foundation (NSF) [GrantCMMI 1824744]. 
}}
\author{Osman Ozaltin}
\author{Reha Uzsoy}
\author{Karl Kempf}
\affil[]{}
\date{}

\begin{document}

\normalsize
\maketitle


\begin{abstract}
    We study the management of product transitions in a semiconductor manufacturing firm that requires the coordination of resource allocation decisions by multiple, autonomous Product Divisions using  a multi-follower bilevel model to capture the hierarchical and decentralized nature of this decision process. Corporate management, acting as the leader, seeks to maximize the firm's total profit over a finite horizon. The followers consist of multiple Product Divisions that must share manufacturing and engineering resources to develop, produce and sell products in the market. Each Product Division needs engineering capacity to develop new products, and factory capacity to produce products for sale while also producing the prototypes and samples needed for the product development process. We model this interdependency between Product Divisions as a generalized Nash equilibrium problem at the lower level and propose a reformulation where Corporate Management acts as the leader to coordinate the resource allocation decisions. We then derive an equivalent single-level reformulation and develop a cut-and-column generation algorithm. Extensive computational experiments evaluate the performance of the algorithm and provide managerial insights on how key parameters and the distribution of decision authority affect system performance.
\end{abstract}
\newpage 


Effective management of \emph{product transitions}, the process of introducing new products into the supply chain to replace older generations, allows companies to maintain and expand market share by incorporating technical advances that satisfy changing customer demand and respond to competitors' offerings \citep{KocaEtal2010}.
Product transitions are particularly challenging for high-tech companies serving multiple markets with short product life cycles and high demand volatility \cite{wu2010improving}. Managing product transitions in these companies requires information sharing and coordinated actions across multiple units to ensure timely access to factory capacity, skilled staff and engineering design resources. Poor decisions can result in failed product transitions with severe business consequences \citep{ErhunConcalves2007}. 

A key feature of product transitions in large high-tech companies is the decentralized decision-making environment. These firms are organized as multiple functional groups, each with its own private information, objective, and operational constraints, and considerable autonomy to address its decision problems \citep{BansalEtal2020}. Therefore, treating the firm as a monolithic entity making centralized decisions with complete information is neither realistic nor desirable. A growing body of research has addressed aspects of product transitions in a decentralized environment, including capacity planning under uncertainty \citep{KarabukWu2003}, supply chain coordination \citep{MallikHarker2004}, capacity coordination and allocation \citep{KarabukWu2005, BansalEtal2020,BansalEtal2022}, and the role of capacity conversion on the equipment renewal and upgrade schedule \citep{LiEtal2014}. Most of this work, however, ignores the hierarchical nature of the decision-making environment, in which corporate management takes strategic and tactical resource allocation decisions that drive the operational decisions of the lower-level functional groups \citep{KhorramfarEtal2022}. The hierarchical nature of decisions mandates a clear line of authority and control, ignoring which can render a decision model impractical and misleading.

In many industries such as semiconductor manufacturing, the distribution of domain knowledge among different functional groups and lines of authority governing the decision-making environment require extensive coordination across the firm \citep{BansalEtal2020}. Although the specifics of product development and high-volume production can differ from one company to another, two principal types of units interact to accomplish a successful product transition: 1) Corporate Management (CORP), which guides the firm toward its strategic business goals by allocating critical resources including budget, factory capacity, and skilled staff; and 2) Product Divisions (PDs) that manage all activities related to new product development and sales for their respective market segments. Each PD is responsible for serving a specific market segment with potentially multiple product lines. The PDs are organized as profit centers within the firm and use their allocated budget to obtain resources, specifically factory capacity and engineering staff, to conduct their operations. Each PD determines the specifications of new products for the market segment it serves and develops a detailed design for each product that outlines its features, the timeline for its introduction, projected demand, and estimated costs \citep{RashKempf2012}. The design step is carried out by engineering staff and is followed by prototype fabrication to identify potential failures and evaluate the manufacturability of the product. Hence, each PD must secure sufficient factory capacity to produce both its prototypes and its products currently being sold in the market. The factory capacity used for prototype fabrication reduces the capacity available for manufacturing products to meet demand, but some allocation is necessary to ensure an ongoing pipeline of new products in the future. 
CORP's objective is to maintain the long-term profitability of the firm, but it can only influence PDs' decisions through the allocation of critical resources among PDs. This substantial decision autonomy of PDs creates an environment where CORP does not have direct knowledge of each PD's operational constraints or the manner in which the PDs compete for factory and engineering resources.

The combination of a complex decision environment and autonomous units involved in the product transition creates an extremely challenging problem. A simpler problem faced by PDs, even in its complete deterministic form, results in a Generalized Nash Equilibrium Problem (GNEP) which is computationally challenging even if the problem associated with each player is a linear program (LP) \citep{FacchineiKanzow2010}. In this paper, we propose a bilevel model to examine the decentralized hierarchical environment for product transitions described above. The leader, CORP, seeks to maximize the firm's total profit, given by the difference between total revenues and expenses of all PDs, subject to the decisions made by the follower's problem. The follower's problem is a GNEP where PDs compete for shared factory capacity and engineering resources to minimize their individual cost over a finite planning horizon. Hence, the decisions of each PD depend on those of all other PDs, presenting a  bilevel program with interdependent followers. We addressed a much more restricted version of the problem where only manufacturing and engineering units interact in our previous work \citep{KhorramfarEtal2022}. In this simplified version, CORP specifies a new product release schedule, and manufacturing allocates factory capacity between products manufactured for sale and prototype fabrication, while engineering units manage their internal resources to complete product development subject to the schedule specified by CORP and the factory capacity allocated by manufacturing. In this study, the decision structure is quite different, better reflecting the actual process in the company motivating this work. 

In contrast to the various simplified models of product transitions that address specific aspects, discussed in Section \ref{sec:Literature-Review}, this paper addresses a substantially more complex - and realistic - decision environment with four different types of agents, three of whom - Manufacturing, Product Engineering and the multiple Product Divisions - all interact at the lower level to obtain a Generalized Nash Equilibrium that responds to the decisions of the leader, CORP. We establish an equivalent single-level equivalent formulation of the bilevel problem and propose a cut and column generation algorithm to obtain exact solutions. Extensive computational experiments examine both the capabilities of the proposed solution algorithm and provide managerial insights.


The remainder of the paper is organized as follows. Section~\ref{sec:Literature-Review} briefly reviews the relevant literature on product transitions in high-tech industries, GNEP, and bilevel programming. We formulate the problem in Section~\ref{sec:Model-Formulation}, and develop our solution approach in Section~\ref{sec:Solution-Approach}. Section~\ref{sec:Computational-Experiment} describes the numerical experiments and presents their results and managerial insights. Section~\ref{sec:Conclusion} summarizes our principal findings and proposes several directions for future research. 

\section{Literature Review} \label{sec:Literature-Review}
We start by reviewing relevant work on product transitions and capacity planning in Section~\ref{subsec:production_trans}, followed by the state of knowledge related to GNEP and bilevel programming is then reviewed in Sections ~\ref{subsec:gnep} and \ref{subsec:bilevel_progam}, respectively. 

\subsection{Product Transitions and Capacity Planning}\label{subsec:production_trans}
Various aspects of product transitions have been studied in recent decades \citep{FerrerSwaminathan2006,GokpinarEtal2010,LeonEtal2011}. \cite{WuEtal2009} examine the impact of initial investment on the time to market, quality improvement, and pricing of new products. Inventory planning decisions without replenishment are studied in \cite{LiEtal2010} considering demand uncertainty and product substitution. \cite{SundaramoorthyEtal2012} address capacity planning in the pharmaceutical industry subject to uncertain outcomes of clinical trials whose failure may prevent a product being brought to market. \cite{DruehlEtal2009} study the timing of new product introduction and identify factors that impact the pace of product transition. \cite{LiangEtal2014} and \cite{LobelEtal2015optimizing} analyze optimal product launch policies under different operational settings in the presence of strategic customers, while \cite{CuiWu2017} examine the impact of customer involvement in the success of new product development. Other studies address the impact of pricing on product transitions, including interactions among timing, pricing, and product design \citep{KocaEtal2010,LiEtal2010}.

Decentralized capacity planning has been studied in a number of papers. \cite{MallikHarker2004} examine capacity allocation among product lines in a setting where competing product line managers request production capacity from a manufacturing unit. Incorporating demand and yield uncertainties using discrete scenarios, \cite{KarabukWu2003} model the decision problems of marketing and production units. They impose constraints forcing the capacity allocation and expansion decisions to be consistent and apply different recourse approximation schemes to decentralize the capacity planning process. A subsequent paper \citep{KarabukWu2005} considers corporate management and production units to develop a game-theoretical model to elicit private information from the production unit to maximize expected corporate profit. \cite{BansalEtal2020} propose two iterative combinatorial auction schemes for a simpler version of the problem considered in this paper to coordinate the negotiation over factory capacity between manufacturing, acting as the auctioneer, and product development units acting as bidders. A subsequent paper \citep{BansalEtal2022} uses mixed-integer programming duality to design a price-based coordination procedure. 

The closest model to the problem we consider in this paper is that of \cite{KhorramfarEtal2022}, which developed a bilevel programming model with two followers for product transition and capacity coordination. Motivated by the semiconductor industry, the model treats corporate management as the leader, and the Manufacturing and Product Engineering groups as followers with the principal negotiation taking place over the allocation of manufacturing capacity between development and production for sale, effectively aggregating all product divisions within Corporate Management. This paper differs in that: 1) it considers the financial aspect of product transition, namely budget allocation, which is the prevalent approach to coordinating activities within multidivisional firms; 2) it models a more complex decision environment involving Product Engineering, Manufacturing and multiple Product Divisions, and thus applicable to industries other than semiconductor manufacturing such as pharmaceutical companies; 3) it recognizes the importance of engineering resources as well as manufacturing capacity, considering the allocation of both engineering resources and manufacturing capacity among the competing PDs. 



\subsection{Generalized Nash Equilibrium Problems}\label{subsec:gnep}
The Generalized Nash Equilibrium Problem (GNEP) is a non-cooperative game in which the strategy of each player depends on those of rival players. A collection of strategies called an \textit{equilibrium; plural, equilibria} is attained when no player can benefit from a unilateral change of strategy. Viewed as a mathematical optimization model, the GNEP is a decentralized model whose decision-makers share a set of constraints known as \textit{coupling constraints}. It has been fruitfully applied to many optimization problems \citep{FacchineiKanzow2010}, including transportation \citep{SagratellaSchmidt2020} 
, electricity markets~\citep{Wang2Etal021,ZhaoEtal2023}, and telecommunication networks~\citep{AltmanWynter2004}.
 
Solution methods for GNEP are predominantly limited to problems with players who control only continuous decision variables \citep{FacchineiKanzow2010,AusselSagratella2017}. Although GNEP with integer variables arise naturally in many applications, this literature has remained limited because the presence of integer variables results in non-convex and non-differentiable problems for which no general algorithmic approach is yet available \citep{SagratellaSchmidt2020}. In the first treatment of GNEP with integer variables, \cite{Sagratella2017a} considers a special case of GNEP, generalized potential games, where all players optimize the same (potential) objective function over the aggregated strategy sets of all players.  Under some assumptions, they establish the attainability of equilibria in a GNEP with mixed-integer variables and linear coupling constraints and propose branching methods to enumerate the problem equilibria. \cite{FabianiGrammatico2019} apply this procedure to a transportation problem with mixed-integer variables. More recently, \cite{SagratellaSchmidt2020} introduced a noncooperative fixed charge transportation problem with different cost structures and constraints. They formulate the problem as a mixed-integer GNEP and establish structural properties that allow them to apply algorithms developed for pseudo-Nash equilibrium problems, a concept we discuss later in this paper. 

\subsection{Bilevel Programming}\label{subsec:bilevel_progam}
Bilevel programming (BP) is a nested optimization problem over a feasible region where the solution to the follower's (lower-level) optimization problem defines a subset of the leader's (upper-level) constraints \citep{book:Bard1998}. Under optimistic BP the leader selects the most advantageous of the followers' responses, whereas under pessimistic BP the leader must choose the least favorable. Bilevel programs provide a powerful tool to model hierarchical decision processes with two decision makers who share a set of common resources but pursue different objectives. The leader makes a decision based on the follower's anticipated response, after which the follower responds to the leader's decision \citep{book:Bard1998}. This inherent hierarchy gives BP a wide range of applications including energy markets, transportation, facility location, military defense, machine learning, and healthcare \citep{KleinertEtal2021_survey,OzaltinEtal2018}. 

The most widely studied BP problem is bilevel linear programming (BLP), in which both leader's and follower's problems are linear programs (LPs) \citep{HemmatiSmith2016}. Although BLPs are non-convex and strongly NP-hard, the optimality of the follower's decisions can be enforced through the Karush-Kuhn-Tucker (KKT) or strong duality conditions of the follower's problem ~\citep{BolusaniRalph2022}. The resulting single-level nonlinear problem can be solved by a combination of branch-and-bound and penalty methods \citep{KleinertEtal2021_survey}.

\cite{BardMoore1990} offer the first algorithmic approach to mixed-integer bilevel linear programs (MIBLP), BPs with integer decision variables in at least one level. They identify node fathoming rules and develop a simple branch-and-bound method. Algorithmic development for the general MIBLP then ceased for nearly two decades until \cite{DenegreRalph2009} proposed a branch-and-cut algorithm, stimulating renewed activity in this field \citep{BolusaniRalph2022}. \cite{XuWang2014} propose a branch-and-bound algorithm with multi-way branching on the slack variables. A subsequent paper \citep{WangXu2017} presents a similar approach that eliminates bilevel infeasible solutions by multi-way branching disjunctions. \cite{LozanoSmith2017} propose an algorithm based on the single-level value function reformulation. \cite{FischettiEtal2018} suggests the use of intersection cuts in a branch-and-cut algorithm. \cite{YueEtal2019} propose a projection-based single-level reformulation that implicitly enumerates the follower's integer variables. Finally, \cite{BolusaniRalph2022} utilizes MILP duality theory to develop a Benders-like decomposition algorithm that approximates the follower's value function. MIBLP has successfully applied to several optimization problems \citep{KleinertEtal2021_survey,Dempe2018}. \cite{HemmatiSmith2016} consider a competitive prioritized set covering a problem that arises in new product development and introduction in a competitive market; they formulate the problem as an MIBLP and develop a cutting-plane algorithm. Other applications of MIBLP include facility location~\citep{BeresnevMelnikov2018}, transportation~\citep{ArslanEtal2018}, healthcare \citep{OzaltinEtal2018}, and  electricity markets \citep{LavigneSavard2000}.

While most of the BP literature considers only a single follower, some studies consider lower-level problems with more than one follower. \cite{ShiEtal2007} study a multi-follower BLP with partially shared variables among followers. \cite{CalveteEtal2019} formulate a rank pricing problem as a nonlinear bilevel program with multiple independent followers and develop a tailored branch-and-cut algorithm. \cite{tavasliouglu2019solving} apply a generalized value function, whose arguments are the objective function gradient and right-hand-side of constraints, to a multiple-follower MIBLP whose followers share a common constraint matrix. \cite{BasilicoEtal2020} study a BLP whose multiple followers play a Nash game, establishing several structural properties and proposing a branch-and-bound method. 
To the best of our knowledge, the only other study of MIBLP with multiple followers playing a GNEP is our own prior work \citep{KhorramfarEtal2022} that considers a substantially simpler decision structure than that addressed in this paper.

\section{Model formulation}\label{sec:Model-Formulation}
\begin{figure}
    \centering
    \includegraphics[width=0.75\textwidth]{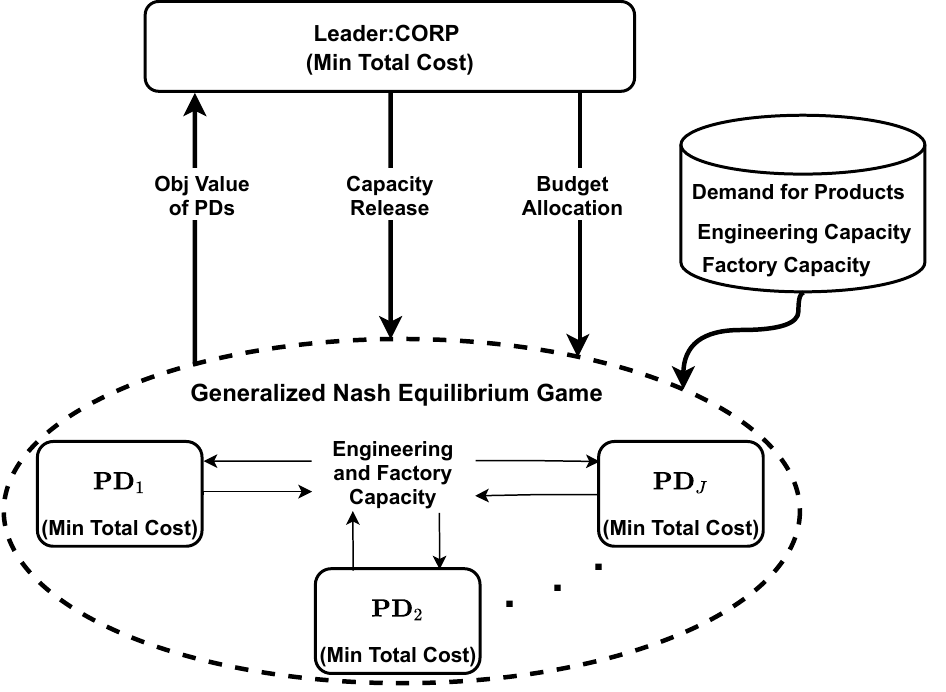}
    \caption{The information flow among decision units in the proposed bilevel model. Given upper-level budget allocation and capacity decisions, PDs compete for factory and engineering capacity as depicted by the solid arrows in the lower level.
Once PDs reach an equilibrium, they communicate their objective function values to the leader. }
    \label{fig:Model-Diag}
\end{figure}

We formulate the product transition problem as a bilevel program to capture its hierarchical decentralized structure. The upper-level problem represents the decisions of Corporate Management (CORP) whose objective is to maximize the total contribution (total revenue minus the total variable cost) of all PDs. CORP allocates budget to PDs, and specifies the allocation of factory capacity between production for sale and engineering in each time period. The PDs use their budget allocation to acquire factory capacity for their product development and manufacturing operations. The lower-level problem is a generalized Nash game played among multiple PDs whose coupling constraints ensure that the usage of factory and engineering capacity are consistent with CORP's decisions. Each PD seeks to minimize its total cost, which includes backorder, production, and inventory holding costs.  Figure~\ref{fig:Model-Diag} illustrates the flow of information in the proposed bilevel model.

\begin{table}[h]   
    \centering
    \caption{Parameters of the bilevel product transition model  for each PD$_j$}
    \label{tab:params-PD}
    \begin{tabular}{l l}
    \toprule
    $D_{jnt}$ & Demand for product $n$ in period $t$\\
    $\pi_{jnt}$ & Unit revenue of product $n$ in period $t$\\
    $r_{jnt}$&  Unit production cost of product $n$ in period $t$\\
    $b_{jnt}$&  Unit backorder cost of product $n$ in period $t$\\
    $H_{jp}^f$ & Factory capacity required for prototype fabrication of new product $p$\\
    $H_{jp}^e$ & Engineering capacity required for new product $p$\\
    $h_{jnt}$& Unit inventory holding cost for product $n$ in period $t$\\
    $C^f_{t}$ & Total factory capacity available in period $t$\\
    $C_{t}^e$ & Total engineering capacity available in period $t$\\
    $c_{j}^f$ & Factory capacity cost \\
    $c_{j}^e$  & Engineering capacity cost \\
         \bottomrule
    \end{tabular}
\end{table}

\begin{table}[h]   
    \centering
    \caption{Decision variables of the bilevel product transition model for each PD$_j$}
    \label{tab:vars-PD}
    \begin{tabular}{l l}
    \toprule
    $X_{jnt} \in \mathbb{Z}^+$ & Factory capacity used for manufacturing of product $n$ of PD $j$ for sale in period $t$ \\
    $V_{jnt} \in \mathbb{Z}^+$ & Backorder of product $n$ of PD $j$ in period $t$ \\
    $I_{jnt} \in \mathbb{Z}^+$ & Inventory level of product $n$ of PD $j$ in period $t$ \\
    $R^f_{jt}\in \mathbb{Z}^+$ & Factory capacity allocated to PD $j$ for product development in period $t$ \\
    $R^e_{jt} \in \mathbb{Z}^+$ & Engineering capacity allocated to PD $j$ in period $t$ \\
    $Z_{jpt}\in \{0,1\}$ & 1 if product $p$ has completed all development activities in or prior to period $t$, 0 otherwise\\
         \bottomrule
    \end{tabular}
\end{table}

Let $\mathcal{T} = \{1,\ldots,T\}$ denote the set of time periods, and $\mathfrak{B}$ the firm's total available budget over the planning horizon. We denote the set of PDs by $\mathcal{J} = \{1,\ldots,J\}$, the set of all current and new products by $N$, and the set of new products currently under development by $P \subset N$. We define $N_j$ and $P_j$ as the sets of all products and new products of $\text{PD}_j$, respectively. Model parameters and variables are defined in Tables ~\ref{tab:params-PD} and \ref{tab:vars-PD}, respectively. The integer variable $\mathcal{B}_j$ represents CORP's budget allocation to $\text{PD}_j$ for the entire planning horizon. Similarly, the integer variables $S_t^f$ and $S_t^e$ represent the amount of factory capacity and engineering resources allocated by CORP to production for sale and product development activities, respectively, in each period $t$. We use boldface symbols to denote vectors of variables; for example, $\boldsymbol{\mathcal{B}}$ denotes the vector of decision variables $\mathcal{B}_j$ for all $j \in \mathcal{J}$. 

Given the upper-level budget allocation $\boldsymbol{\mathcal{B}}$ and capacity decisions $\boldsymbol{S} =(\boldsymbol{S}^f,\boldsymbol{S}^e)$, we use GNEP($\boldsymbol{\mathcal{B}},\boldsymbol{S}$) to denote the generalized Nash equilibrium game as well as its equilibria in the lower level. 
The outcome of GNEP($\boldsymbol{\mathcal{B}},\boldsymbol{S}$) is an equilibrium in which each PD incurs a cost denoted by $\phi_j$. This product transition model can be formulated as follows:
\begin{subequations}\label{model:CORP}
    \begin{align}
       \hspace{-1cm} \text{[PTM-Nash]} \  \max \ & \ \theta = \sum_{j \in \mathcal{J}} \sum_{n \in N^j}\sum_{t \in \mathcal{T}} \pi_{jnt}(D_{jnt}-V_{jnt}+V_{jn,t-1})-\sum_{j \in \mathcal{J}} \phi_j \label{M-CORP:obj}\\
        \text{s.t.} \ &\sum_{j \in \mathcal{J}} \mathcal{B}_{j} = \mathfrak{B} & \label{M-CORP:c1}\\
        &S^f_{t} \leq C^f_{t}& t\in \mathcal{T}  \label{M-CORP:c2}\\
        & S^e_{t} \leq C^e_{t} & t\in \mathcal{T}\label{M-CORP:c3}\\
        & \boldsymbol{\phi}  \in \text{GNEP} (\boldsymbol{\mathcal{B}},\boldsymbol{S})& \label{M-CORP:c4}\\
        &\mathcal{B}_j, S^f_{t}, S^e_{t} \in \mathbb{Z}^+  & j \in \mathcal{J}, t\in \mathcal{T}.
        \label{M-CORP:c5}
    \end{align}
\end{subequations}

The objective function \eqref{M-CORP:obj} maximizes the total contribution, given by the difference between the total revenue and the total cost of all PDs. Constraint~\eqref{M-CORP:c1} distributes the total available budget to PDs, while \eqref{M-CORP:c2} and \eqref{M-CORP:c3} prevent factory capacity allocation $S^f_t$ and engineering capacity allocation $C^e_t$ from exceeding the available capacities $C^f_t$ and $C^e_t$. The generalized Nash game among PDs in the lower level is represented by~\eqref{M-CORP:c4}. 
PTM-Nash is an optimistic bilevel model which allows the leader, CORP, to select the most favorable lower-level equilibrium~\citep{KleinertEtal2021_survey}.

GNEP($\boldsymbol{\mathcal{B}},\boldsymbol{S}$) involves a set of interacting PD problems with the same constraint structure and objective function but different parameter values. 
Each PD$_j$ problem in GNEP($\boldsymbol{\mathcal{B}},\boldsymbol{S}$) can be modeled as follows:
\begin{subequations}\label{Model:PD}
    \begin{align}
   \hspace{-1cm}[\text{PD}_j (\boldsymbol{\mathcal{B}},\boldsymbol{S})] \quad   \min \ & \phi_j = \sum_{n \in N^j}\sum_{t \in \mathcal{T}} [ b_{jnt}V_{jnt}+   r_{jnt}X_{jnt} + h_{jnt}I_{jnt} ]\hspace{-3cm}&\label{M-PD:obj} \\
       \text{s.t.} \ &\sum_{i \in \mathcal{J}} R^f_{it}= S^f_{t} & \hspace{-1.5cm} t \in \mathcal{T}\label{M-PD:c1}\\
        &\sum_{i \in \mathcal{J}} R^{e}_{it} = S^e_{t} & \hspace{-1.5cm}t \in \mathcal{T}\label{M-PD:c2}\\
        & \sum_{t \in \mathcal{T}}R^f_{jt} c_{jt}^f+ \sum_{t \in \mathcal{T}}R^e_{jt} c_{jt}^{e} \leq \mathcal{B}_j  &\hspace{-1.5cm}\label{M-PD:c3}\\
        &\sum_{p \in P^j}H^f_{jp}Z_{jpt}+\sum_{n \in N^j}X_{jnt} \leq  R^f_{jt} &\hspace{-1.5cm} t \in \mathcal{T}\label{M-PD:c4}\\
        &\sum_{p \in P^j}H^e_{jp}Z_{jpt} \leq  R^e_{jt} & \hspace{-1.5cm}t \in \mathcal{T}\label{M-PD:c5}\\
        &I_{jnt} = I_{jn,t-1}+X_{jnt} -(D_{jnt} +V_{jn,t-1} -V_{jnt}) &\hspace{-1.5cm} n \in N_j, t \in \mathcal{T} \label{M-PD:c6} \\ 
         & \sum_{t \in \mathcal{T}} Z_{jpt} \leq 1 &\hspace{-1.5cm} p \in P_j\label{M-PD:c7} \\
        &X_{jpt} \leq C^f_{t}\sum_{\tau \leq t} Z_{jp\tau}&\hspace{-1.5cm} p \in P^j, \ t \in \mathcal{T}\label{M-PD:c8} \\
         &I_{jnt},X_{jnt},V_{jnt},R^e_{jt},R^f_{jt} \in \mathbb{Z}^+, Z_{jpt} \in \{0,1\} &\hspace{-1.3cm} \forall t\in \mathcal{T}, n\in N^j, p \in P^j \label{M-PD:c9}
    \end{align}
\end{subequations}
 
The objective function~\eqref{M-PD:obj} minimizes the sum of backorder, production, and inventory holding costs over the planning horizon. The \textit{coupling constraints}~\eqref{M-PD:c1} and \eqref{M-PD:c2} capture the factory and engineering capacity usage by all PDs in each period. Note that the feasible region of each PD depends on both the upper-level decisions and the decisions of other PDs due to the coupling constraints. Constraint~\eqref{M-PD:c3} limits the total cost of the factory  and engineering capacity the PD can use over the planning horizon to the budget $\mathcal{B}_j$ allocated by CORP to PD$_j$. Constraints~\eqref{M-PD:c4} ensure that the factory capacity used by PD$_j$ for product development($\sum_{p \in P^j}H^f_{jp}Z_{jpt}$) and for production for sale ($\sum_{n \in N^j}X_{jnt}$) do not exceed the acquired factory capacity $R^f_{jt}$. Similarly, constraints~\eqref{M-PD:c5} restrict PD$j$'s engineering capacity usage for new product development in each time period. Constraints~\eqref{M-PD:c6} are inventory balance equations across time periods. Constraints~\eqref{M-PD:c7} ensure that each new product consumes capacity for product development only once in the planning horizon, i.e., all development activities for a product can be completed in a single period.  This assumption is easily relaxed by considering multiple product development stages \citep{BansalEtal2022}; the proposed bilevel framework remains applicable. Finally, constraints~\eqref{M-PD:c8} ensure that new products can only be manufactured after development has been completed. We refer to constraints~\eqref{M-PD:c1}-\eqref{M-PD:c9} as \texttt{PD$_j$-feasibility}$(\boldsymbol{\mathcal{B}},\boldsymbol{S})$. We also define \texttt{PD$_j$-consts}$(\boldsymbol{\mathcal{B}},\boldsymbol{R})$, where $\boldsymbol{R}= (\boldsymbol{R}^f,\boldsymbol{R}^e)$, to refer to constraints~\eqref{M-PD:c3}-\eqref{M-PD:c9} with specified values of the budget, factory capacity, and engineering capacity in each period.

\section{Solution Approach} \label{sec:Solution-Approach}
After providing some technical background about pseudo-Nash equilibrium games in Section~\ref{subsec:pseudo-Nash}, we present a single-level reformulation of the PTM-Nash problem in Section~\ref{subsec:single-lvel-reform} and develop a solution algorithm in Section~\ref{subsec:ccg algorithm}.

\subsection{Pseudo-Nash Equilibrium Games}\label{subsec:pseudo-Nash}
Let $\mathcal{P}$ be the set of players in a generalized Nash equilibrium game, $x^\nu$ the set of variables controlled by player $\nu \in \mathcal{P}$ and $x^{-\nu}$ those controlled by other players $\gamma \in \mathcal{P}\setminus \{\nu\}$. Furthermore, let $K_\nu(x^{-\nu})$ and $S_\nu(x^{-\nu})$ denote the strategy (i.e., feasible solution) and response (i.e., optimal solution) sets of player  $\nu$, respectively. A solution $\hat{x}=(\hat{x}^\nu,\hat{x}^{-\nu})$ is a generalized Nash equilibrium if and only if it satisfies $\hat{x}^\nu \in S_\nu(\hat{x}^{-\nu})$ for all  $\nu \in \mathcal{P}$. In other words, $\hat{x} \in S(\hat{x}) \equiv \Pi_{\nu\in \mathcal{P}} S_\nu(\hat{x}^{-\nu})$. Any generalized Nash equilibrium $\hat{x}=(\hat{x}^\nu,\hat{x}^{-\nu})$ satisfies $\hat{x}^\nu \in K_\nu(\hat{x}^{-\nu})$ for all  $\nu \in \mathcal{P}$ since $S_\nu(x^{-\nu}) \subseteq K_\nu(x^{-\nu})$. That is, $\hat{x} \in K(\hat{x}) \equiv \Pi_{\nu\in \mathcal{P}} K_\nu(\hat{x}^{-\nu})$. We denote by $FP(S) \equiv \{x: x \in S(x)\}$ the set of fixed points of the point-to-set mapping $S$ that defines the set of all generalized Nash equilibrium solutions. Similarly, we denote by $FP(K) \equiv \{x: x \in K(x)\}$ the set of fixed points of the point-to-set mapping $K$.



\begin{definition}
\label{Def:pseudo-Nash}
 \citep{AusselSagratella2017} A generalized Nash equilibrium game is a \textit{pseudo-Nash equilibrium game} if its set of equilibrium solutions $FP(S) \equiv \cup_{x \in K(x)}S(x)$.
\end{definition}
Based on Definition~\ref{Def:pseudo-Nash}, an equilibrium of a pseudo-Nash game can be computed by identifying a solution $\hat{x} \in K(\hat{x})$ and then finding the response set $S_\nu(\hat{x}^{-\nu})$ of each player $\nu \in \mathcal{P}$. Note that $S_\nu(\hat{x}^{-\nu})$ does not depend on the decisions of other players  once $\hat{x}^{-\nu}$ is fixed.  \cite{AusselSagratella2017} have shown that any generalized Nash equilibrium game with linear objective functions and linear constraints whose coupling constraints are exclusively equality constraints is a pseudo-Nash equilibrium game.


\begin{proposition}\label{prop:gnep_is_pseudoNash}
$\text{GNEP}(\boldsymbol{\mathcal{B}},\boldsymbol{S})$ is a pseudo-Nash equilibrium game.
\end{proposition}
\noindent \textit{Proof.}  Each PD in $\text{GNEP}(\boldsymbol{\mathcal{B}},\boldsymbol{S})$  has a  linear objective function and linear constraints. Moreover, coupling constraints \eqref{M-PD:c1} and \eqref{M-PD:c2} are exclusively equality constraints.
\halmos

This proposition establishes that an equilibrium can be obtained from a feasible solution to the lower-level problem. Given budget allocation $\boldsymbol{\mathcal{B}}$ and capacity decisions $\boldsymbol{S}$, consider the following \textit{equilibrium problem (EP)} whose objective function is the sum of the objective functions of the individual PDs, subject to all constraints in the lower level problem. 

\begin{subequations}
    \begin{align*}
      [\text{EP}(\boldsymbol{\mathcal{B}},\boldsymbol{S})]  \min \ & \sum_{j\in \mathcal{J}} \sum_{n \in N^j}\sum_{t \in \mathcal{T}} [b_{jnt}V_{jnt}+   r_{jnt}X_{jnt} + h_{jnt}I_{jnt}]\hspace{-3cm}&\\
         s.t. \ & \texttt{PD$_j$-feasibility}(\boldsymbol{\mathcal{B}},\boldsymbol{S}) & j \in \mathcal{J}\\
    \end{align*}
\end{subequations}

The solution of $\text{EP}(\boldsymbol{\mathcal{B}},\boldsymbol{S})$ is an equilibrium of GNEP($\boldsymbol{\mathcal{B}},\boldsymbol{S}$) from Proposition~\ref{prop:gnep_is_pseudoNash}. However, this solution may not be an optimistic response to the leader because it does not include the revenue term in the CORP's objective function~\eqref{M-CORP:obj}. 

\begin{proposition}\label{prop:multiple-equilibrium}
Given budget allocation $\boldsymbol{\mathcal{\hat{B}}}$ and capacity decisions $\boldsymbol{\hat{S}}$, there exists an equilibrium of GNEP($\boldsymbol{\mathcal{\hat{B}}},\boldsymbol{\hat{S}}$)  corresponding to any nonnegative $\boldsymbol{\hat{R}}=(\boldsymbol{\hat{R}^f},\boldsymbol{\hat{R}^e})$ such that:
\begin{align*}
     &\sum_{i \in \mathcal{J}} \hat{R}^f_{it}= \hat{S}^f_{t} &t \in \mathcal{T}\\
        &\sum_{i \in \mathcal{J}} \hat{R}^{e}_{it} = \hat{S}^e_{t} & t \in \mathcal{T}\\
\end{align*}
\end{proposition}
\noindent \textit{Proof.} 
Let $\sum_{i' \in \mathcal{J} \backslash i} \hat{R}^f_{i't}$ and $\sum_{i' \in \mathcal{J} \backslash i} \hat{R}^e_{i't}$ be the summation of optimal factory and engineering capacity acquisitions for all players except $i$, respectively. Then ${R}^f_{it}= S^f_{t}-\sum_{i' \in \mathcal{J} \backslash i} \hat{R}^f_{i't}=\hat{R}^f_{it}$ in any feasible strategy of player $i$. Therefore, $\hat{\mathcal{X}}=(\hat{I},\hat{X},\hat{V},\hat{R^e},\hat{R^f})$ is an equilibrium of GNEP$(\boldsymbol{\hat{B}} , \boldsymbol{\hat{S}})$.
\halmos

Proposition~\ref{prop:multiple-equilibrium} implies that the leader can choose an equilibrium by specifying the factory and engineering capacity allocation for each PD$_j$ in each period $t$. In other words, setting the values for $\boldsymbol{\hat{R}}$ allows the optimistic leader, CORP, to select its most preferred equilibrium of the lower-level problem.
We use this fact to formulate the following bilevel problem:

\begin{subequations}\label{model:ptm}
    \begin{align}
       \hspace{-1.2cm} [\text{PTM}] \quad \max \ & \ \theta = \sum_{j \in \mathcal{J}} \sum_{n \in N^j}\sum_{t \in \mathcal{T}} \pi_{jnt}(D_{jnt}-V_{jnt}+V_{jn,t-1}) - \sum_{j \in \mathcal{J}} \phi_j\label{M-PTM:obj}\\
        \text{s.t.} \ &\sum_{j \in \mathcal{J}} \mathcal{B}_{j} \leq \mathfrak{B} &\label{M-PTM:c1}\\
        &\sum_{i \in \mathcal{J}} R^f_{it}\leq C^f_{t} &t \in \mathcal{T}\label{M-PTM:c2}\\
        &\sum_{i \in \mathcal{J}} R^{e}_{it} \leq C^e_{t} & t \in \mathcal{T}\label{M-PTM:c3}\\
        &\phi_j\in \argmin \texttt{PD$_j$-optimality}(\boldsymbol{\mathcal{B}},\boldsymbol{R}) & j \in \mathcal{J},\label{M-PTM:c4}
    \end{align}
\end{subequations}
where $R^f_{it}$ and $R^e_{it}$ are now upper-level variables and \texttt{PD$_j$-optimality}($\boldsymbol{\mathcal{B}},\boldsymbol{R}$) is formulated as $min \{\phi_j:\texttt{PD$_j$-consts}(\boldsymbol{\mathcal{B}},\boldsymbol{R})\}$. 

\begin{proposition}\label{prop:equi}
The bilevel problem PTM is equivalent to PTM-Nash.
\end{proposition}
\noindent \textit{Proof.} 
We establish the result by showing that any feasible solution to PTM corresponds to a feasible solution to PTM-Nash and vice versa. Let $(\boldsymbol{\mathcal{\hat{B}}},\boldsymbol{\hat{R}})$ a feasible solution to PTM. define $\boldsymbol{\hat{S}}$ as
\begin{align*}
    &\hat{S}^f_t = \sum_{j \in \mathcal{J}} \hat{R}^f_{it} & t \in \mathcal{T}\\
    &\hat{S}^e_t = \sum_{j \in \mathcal{J}} \hat{R}^e_{it} & t \in \mathcal{T}
\end{align*}
$(\boldsymbol{\mathcal{\hat{B}}},\boldsymbol{\hat{S}})$ satisfy constraints~\eqref{M-CORP:c1}-\eqref{M-CORP:c3}, and the optimality conditions of each PD problem by ~\eqref{M-PTM:c4}. Note that
\begin{align*}
&\hat{R}^f_{it}=\hat{S}^f_t-\sum_{j \in \mathcal{J}\backslash i}\hat{R}^f_{jt}&t \in \mathcal{T}\\
&\hat{R}^e_{it}=\hat{S}^e_t-\sum_{j \in \mathcal{J}\backslash i}\hat{R}^e_{jt}&t \in \mathcal{T}
\end{align*}
Therefore, if $\hat{\phi}_j\in \argmin \texttt{PD$_j$-optimality}(\boldsymbol{\mathcal{\hat{B}}},\boldsymbol{\hat{R}})$, then $\hat{\phi}_j \in \argmin \text{PD}_j(\boldsymbol{\mathcal{\hat{B}}},\boldsymbol{\hat{S}})$, implying that $(\boldsymbol{\mathcal{\hat{B}}},\boldsymbol{\hat{S}})$ is an  equilibrium of $\text{GNEP}(\boldsymbol{\mathcal{\hat{B}}},\boldsymbol{\hat{S}})$ and a feasible solution to PTM-Nash.

Now let $(\boldsymbol{\mathcal{B}'},\boldsymbol{S}')$ be an upper-level and $(\boldsymbol{R}')$ a corresponding lower-level feasible solution to PTM-Nash. This feasible solution satisfies the upper-level constraints~\eqref{M-PTM:c1}-\eqref{M-PTM:c3}. It also satisfies constraints~\eqref{M-PTM:c4} since each PD's response is optimal in an equilibrium of GNEP($\boldsymbol{\mathcal{B}'},\boldsymbol{S}'$). Note that for every $\text{PD}_j(\boldsymbol{\mathcal{{B}}'},\boldsymbol{{S}}')$ we have
\begin{align*}
&{S}'^f_t=\sum_{j \in \mathcal{J}}R'^f_{jt}&t \in \mathcal{T}\\
&{S}'^e_t=\sum_{j \in \mathcal{J}}R'^e_{jt}&t \in \mathcal{T}
\end{align*}

Therefore, if $\phi'_j \in \argmin \text{PD}_j(\boldsymbol{\mathcal{B}'},\boldsymbol{S}')$, then $\phi'_j\in \argmin \texttt{PD$_j$-optimality}(\boldsymbol{\mathcal{B}'},\boldsymbol{R}')$, and $(\boldsymbol{\mathcal{B}'},\boldsymbol{R}')$ is a feasible solution to PTM. The fact that both models have the same objective function concludes the proof. \halmos



\subsection{Single-level Reformulation} \label{subsec:single-lvel-reform}
We use the result established in Proposition~\ref{prop:equi} to develop an equivalent single-level reformulation of PTM-Nash through enumeration of all possible budget allocations. 
Let $\mathcal{G}$ denote the set of all possible distinct budget allocations by  CORP to the $J$ number of PDs, and $G$ be its index set. Then $\mathcal{G}$ is equivalent to the weak composition of $\mathfrak{B}$ into $J$ nonnegative integers \citep{book:Sagan2020} with cardinality
$$|\mathcal{G}| = \binom{\mathfrak{B}+J-1}{J-1}.$$

$|\mathcal{G}|$ is finite and depends on the number of PDs and the total available budget $\mathfrak{B}$.
The leader CORP has at least one optimal capacity allocation $\boldsymbol{\hat{R}}_g =(\boldsymbol{\hat{R}}^f,\boldsymbol{\hat{R}}^e)$ corresponding to each budget allocation $\boldsymbol{\mathcal{\hat{B}}}_g$ in $g \in {G}$. Given $(\mathcal{B}_g,\boldsymbol{\hat{R}}_g)$, we can solve the $\texttt{PD$_j$-optimality}(\mathcal{B}_g,\boldsymbol{\hat{R}}_g)$ to obtain the corresponding optimal objective function value  $\phi^*_{gj}$ of the PD, since for any  $(\mathcal{B}_g,\boldsymbol{\hat{R}}_g)$  the lower level problem is feasible. Thus, the upper-level solutions can be enumerated over budget allocations.
We obtain the following single-level reformulation of PTM where the optimal response of every PD to each possible budget allocation is enforced through additional constraints:

\begin{subequations}\label{model:MP}
    \begin{align}
       \hspace{-1.2cm} [\text{MP}(\mathcal{G})] \quad \max \ & \ \theta = \sum_{j \in \mathcal{J}} \sum_{n \in N^j}\sum_{t \in \mathcal{T}} \pi_{jnt}(D_{jnt}-V_{jnt}+V_{jn,t-1})-\sum_{j \in \mathcal{J}} \phi_j\label{M-MP:obj}\\
        \text{s.t.} \ &\sum_{j \in \mathcal{J}} \mathcal{B}_{j} \leq \mathfrak{B} &\label{M-MP:c1}\\
        &\sum_{i \in \mathcal{J}} R^f_{it}\leq  C^f_{t} & \hspace{-1.5cm} t \in \mathcal{T}\label{M-MP:c2}\\
        &\sum_{i \in \mathcal{J}} R^{e}_{it} \leq C^e_{t} & \hspace{-1.5cm}t \in \mathcal{T}\label{M-MP:c3}\\
      & \texttt{PD$_j$-const}(\boldsymbol{\mathcal{B}},\boldsymbol{R}) & j \in \mathcal{J}\label{M-MP:c4}\\
       & \phi_j = \sum_{n \in N^j}\sum_{t \in \mathcal{T}} [b_{jnt}V_{jnt}+   r_{jnt}X_{jnt} + h_{jnt}I_{jnt}]  & j \in \mathcal{J}\label{M-MP:c5}\\
         & R^e_{jt} \geq \hat{R}^e_{gjt}(1-\alpha_{gj}) &\hspace{-2cm} g \in G, j\in \mathcal{J}, t\in \mathcal{T}\label{M-MP:c6}\\
         & R^f_{jt} \geq \hat{R}^f_{gjt}(1-\alpha_{gj}) &\hspace{-2cm} g \in G,  j\in \mathcal{J}, t\in \mathcal{T}\label{M-MP:c7}\\
         &\sum_{j \in \mathcal{J}} (|\mathcal{B}_j-\mathcal{\hat{B}}_{gj}|) \geq   \frac{\sum_{j \in \mathcal{J}}\alpha_{gj}}{J}   &g \in  G\label{M-MP:c8}\\
          &\phi_j \leq \phi^{*}_{gj}+M\alpha_{gj} & \hspace{-2cm} g  \in G,j \in \mathcal{J}\label{M-MP:c9}\\
         &\alpha_{gj}\in \{0,1\},\mathcal{B}_j,C^f_{t},C^e_{t} \in \mathbb{Z}^+ &\hspace{-2cm} g\in G, j \in \mathcal{J},t\in \mathcal{T}
    \end{align}
\end{subequations}

The objective function value of PD$_j$ corresponding to budget allocation $g\in G$ is denoted by $\phi^{*}_{gj}$.
The mixed-integer program MP($\mathcal{G}$) determines the values of the upper and lower-level variables in PTM-Nash. Lower level feasibility is ensured by the constraint sets $\texttt{PD$_j$-const}(\boldsymbol{\mathcal{B}},\boldsymbol{R})$ for $j \in \mathcal{J}$,and optimality of each PD's decisions by ~\eqref{M-MP:c6}-\eqref{M-MP:c9}. Binary variable $\alpha_{gj}$ = 0 if PD$_j$ has access to the sufficient factory and engineering capacity under budget allocation $g$. In other words, if $\alpha_{gj}=0$, both constraints~\eqref{M-MP:c6} and \eqref{M-MP:c7} are binding, and the allocated capacities in every time period $t$ must exceed what PD$_j$ requires to achieve the objective value of $\phi^*_{gj}$. 
If enough capacity is allocated in every time period, constraint~\eqref{M-MP:c9} ensures that  $\phi_j$ does not exceed the optimal objective value of PD$_j$, i.e. $\phi_j \leq \phi^{*}_{gj}$, enforcing the optimality of PD$_j$'s decisions.
On the other hand, if sufficient capacity is not allocated in every time period, then $\alpha_{gj} = 1$ and constraints~\eqref{M-MP:c6} and \eqref{M-MP:c7} are not binding. In this case, constraint~\eqref{M-MP:c8} ensures the generation of a new budget allocation. 
More specifically, the right-hand side of ~\eqref{M-MP:c8} takes a positive value between 0 and 1, forcing the absolute difference of the budget allocation $\mathcal{B}_j$ and allocation plan $\mathcal{\hat{B}}_{gj}$ to be at least 1. This results in new, distinct budget allocations.

\begin{proposition}
The single-level model MP($\mathcal{G}$) is equivalent to PTM-Nash
\end{proposition}
\noindent \textit{Proof.} We establish the equivalency by showing that both models have the same feasible regions and objective functions. Let $\boldsymbol{\mathcal{{X}}}$ represent the set of all decision variables of PTM-Nash, and let $\boldsymbol{\mathcal{\bar{X}}}$ be a feasible solution. Define
\begin{align*}
    \bar{\alpha}_{gj} =
    \begin{cases}
    & 0 \text{ if } R^f_{jt} \geq \hat{R}^f_{gjt} \text{ and } R^e_{jt} \geq \hat{R}^e_{gjt}\\
    & 1 \text{ otherwise}
    \end{cases}
\end{align*}
Then $(\boldsymbol{\mathcal{\bar{X}}},\bar{\alpha})$ satisfies constraints~\eqref{M-MP:c6} and \eqref{M-MP:c7} by construction. It also satisfies \eqref{M-MP:c9} as $\bar{\phi}_{gj}\in \argmin \texttt{PD$_j$-optimality}(\boldsymbol{\mathcal{B}},\boldsymbol{R})$, and is thus a feasible solution to MP($G$). Now consider a feasible solution $(\boldsymbol{\hat{\mathcal{{X}}}},{\hat{\alpha}})$ to MP($G$). The solution $\boldsymbol{\hat{\mathcal{{X}}}}$ is feasible to upper and lower level constraints of PTM-Nash. The index set of all feasible resource allocations to $\texttt{PD}_j\texttt{-optimality}(\boldsymbol{\hat{\mathcal{B}}},\boldsymbol{\hat{R}})$ is given by  $\hat{G}_j = \{g\in G| R^f_{jt} \geq \hat{R}^f_{gjt} \text{ and }R^e_{jt} \geq \hat{R}^e_{gjt}\}$. Since $\hat{\alpha}_{kj}=0$ for every $k\in \hat{G}_j$ due to constraints~\eqref{M-MP:c6} and \eqref{M-MP:c7}, constraints~\eqref{M-MP:c9} ensure that
$\hat{\phi}_j \leq \phi^{*}_{gj}$, rendering $\boldsymbol{\hat{\mathcal{{X}}}}$ a feasible solution to PTM-Nash. The proof is now complete as both models share the same objective function.  \halmos

\subsection{Cut-and-Column Generation Algorithm}\label{subsec:ccg algorithm}
MP($\mathcal{G}$) reformulates the problem with an additional set of constraints whose size grows exponentially with the number of time periods and PDs, rendering the problem intractable even for small instances. To address this issue we propose a cut-and-column generation (CCG) algorithm that starts with a subset of constraints~\eqref{M-MP:c6}-\eqref{M-MP:c9} and iteratively adds constraints. In each iteration $k$ the algorithm solves a restricted master problem $\text{MP}(\mathcal{G}^k)$ where $\mathcal{G}^k\subseteq \mathcal{G}$ and $\mathcal{G}^0=\emptyset$ whose optimal objective value $\theta^k$ is a lower bound LB$^k$ on the optimal objective value $\theta^*$ as the feasible region of $\text{MP}(\mathcal{G}^k)$ contains that of MP($\mathcal{G}$). Let $\boldsymbol{\mathcal{\bar{X}}}^k$ be the solution of $\text{MP}(\mathcal{G}^k)$ in iteration $k$, $(\boldsymbol{{I}}^k,\boldsymbol{{X}}^k,\boldsymbol{{B}}^k,\boldsymbol{{Z}}^k)$ the solution of $\texttt{PD$_j$-optimality}(\boldsymbol{\bar{\mathcal{B}}}^k, \boldsymbol{\bar{R}}^k)$ and ${\phi}_j^k$ its objective value. Then an upper bound on $\theta^k$ is given by
$$
\text{UB}^k =  \sum_{j \in \mathcal{J}} \sum_{n \in N^j}\sum_{t \in \mathcal{T}} \pi_{jnt}({B}^k_{jnt}-{B}^k_{jn,t-1})+\sum_{j \in \mathcal{J}} {\phi}^k_j
$$
as $\boldsymbol{\mathcal{\bar{X}}}^k$ is a feasible solution for MP($\mathcal{G}$).
If $\text{UB}^k-\text{LB}^k< \epsilon$, where $\epsilon$ is the optimality gap, the upper and lower bounds are sufficiently close for termination. Otherwise we define a new $\alpha_{gj}$ and add a new set of constraints ~\eqref{M-MP:c6}-\eqref{M-MP:c9} to the restricted master problem.  Algorithm~\ref{alg:ccg-PD} presents the CCG algorithm.

\begin{algorithm}
\SetAlgoLined
 \textbf{Initialization:} set $k=0$, $\mathcal{G}^k=\emptyset$\label{alg-init}\\
 \While{$\mathcal{G}^k \subseteq \mathcal{G}$}{
 Solve MP($\mathcal{G}^k$).\label{alg-Solve-rmp}\\
 \If{MP($\mathcal{G}^k$) is infeasible or $\mathcal{G}^k=\mathcal{G}$} {
  MP($\mathcal{G}$) is infeasible, STOP. \label{alg-inf}
 }
  Let $\bar{\mathcal{X}}^k=(\boldsymbol{\bar{\mathcal{B}}}^k,\boldsymbol{\bar{R}}^k,\boldsymbol{\bar{Z}}^k,\boldsymbol{\bar{X}}^k,\boldsymbol{\bar{I}}^k,\boldsymbol{\bar{B}}^k)$ be the optimal solution to MP($\mathcal{G}^k$) and $\bar{\theta}^k$ be its optimal objective value.\\
  Set $\text{LB}^k = \bar{\theta}^k$.\\
   For all $j\in \mathcal{J}$, solve $\texttt{PD$_j$-optimality}(\boldsymbol{\bar{\mathcal{B}}}^k, \boldsymbol{\bar{R}}^k)$, obtain the optimal solution $(\boldsymbol{{I}}^k,\boldsymbol{{X}}^k,\boldsymbol{{B}}^k,\boldsymbol{{Z}}^k)$ and optimal objective value $\phi_j^k$\label{alg-sol}.\\
     Set $\text{UB}^k =  \sum_{j \in \mathcal{J}} \sum_{n \in N^j}\sum_{t \in \mathcal{T}} \pi_{jnt}({B}^k_{jnt}-{B}^k_{jn,t-1})+\sum_{j \in \mathcal{J}} {\phi}^k_j$.\\
  \If{$\text{UB}^k-\text{LB}^k< \epsilon$ \label{alg-cond-opt}}
  {
  Return the optimal bilevel solution $\bar{\mathcal{X}}^k$. 
  }\label{stop1-Line}
\Else{
Add constraints ~\eqref{M-MP:c6} to \eqref{M-MP:c9} to MP($\mathcal{G}^k$). Set 
}
     Set $\mathcal{G}^k = \mathcal{G}^k \cup \{\bar{\mathcal{B}}_j, \bar{R}^f_{jt},\bar{R}^e_{jt}\}$, $k = k+1$.
}
 \caption{Cut and column generation (CCG) algorithm}
 \label{alg:ccg-PD}
\end{algorithm}
\begin{proposition}
Algorithm~\ref{alg:ccg-PD} terminates in finite number of iterations.
\end{proposition}

\noindent \textit{Proof.} At each iteration a new set $\boldsymbol{\mathcal{B}}$ forming a weak composition of $\mathfrak{B}$ is generated. The finiteness of the algorithm follows directly since the number |$\mathcal{G}$| of distinct integer budget allocations is finite. \halmos

\begin{proposition}
Algorithm~\ref{alg:ccg-PD} either returns the optimal solution to MP($\mathcal{G}$) or correctly identifies infeasibility.
\end{proposition}

\noindent \textit{Proof.} 
MP($\mathcal{G}^k$) is a relaxation of MP($\mathcal{G}$), so the infeasibility of the former establishes the infeasibility of the latter, causing the algorithm to terminate in line~\ref{alg-inf}. If the condition in line~\ref{alg-cond-opt} is satisfied, the solution $\bar{\mathcal{X}}^k$ is bilevel optimal. If neither the infeasibility of MP($\mathcal{G}^k$) nor the optimality of $\bar{\mathcal{X}}^k$ can be confirmed, the algorithm will continue generating all feasible budget allocations at which point $\mathcal{G}^k=\mathcal{G}$ and the algorithm will stop at line~\ref{alg-inf} by announcing the infeasibility of the problem. \halmos

\section{Computational Experiments} \label{sec:Computational-Experiment}
In this section, we first describe the generation of the random test instances and the implementation of the proposed algorithm.
We then examine the performance of the proposed solution approach.  We also provide managerial insights by examining the impact of competition and the number of new products developed by each PD.

\subsection{Instance Generation and Implementation Details} \label{sec:instance-generation}
We first generate parameters for all current and new products and then populate the parameters associated with each PD.  
We generate demand from one of the four discrete uniform distributions presented in Table \ref{tab:Par_vals} to account for the fact that high-tech companies with fast product transitions operate in different market segments \citep{KhorramfarEtal2022,RashKempf2012}. We consider a realistic demand pattern observed in many industries where demand for current products diminishes to zero over time while the demand for new products first increases and then stabilizes. Specifically, demand for the current products diminishes to zero after the first 40\% to 70\% of the planning horizon. The demand for new products starts increasing after the first 30\% to 50\% of the planning periods. The values and ranges of parameters in Table~\ref{tab:Par_vals} except the holding costs are based on \citep{KhorramfarEtal2022}.


\begin{table} 
    \centering
    \caption{Demand Distributions Per Period}
    \label{tab:Par_vals}
    \begin{tabular}{ll|ll}
    \toprule
         Demand in Market 1 & U$[2,20]$ & $h$ (holding cost) & $1$ \\
        Demand in Market 2 & U$[15,20]$ & $r$ (production cost) & $2h$\\
         Demand in Market 3 & U$[20,50]$ & $b$ (backorder cost) & $10h$\\
         Demand in Market 4 & U$[30,150]$& $\pi$ (revenue per unit) & U$[20,30]$\\
         \bottomrule
    \end{tabular}
\end{table}

We assume that the factory capacity in each period is uniformly distributed between 70\% and 130\% of the total demand for all products in that period. 
We ensure that the factory capacity is sufficient to meet the demand for all products during the planning horizon on average.
The engineering capacity in each period is randomly generated  from a discrete uniform distribution between 2 and 20 engineers. 

Once the factory and engineering capacities are determined, we generate the values of $H^f_{jp}$ and $H^e_{jp}$, i.e., the factory and engineering capacities required for the development of product $p$ by PD$_j$. 
We first calculate the average factory and engineering capacity available to each PD in each time period by computing their total factory and engineering capacity requirements over the entire planning horizon and dividing these values by the number of periods and PDs. We then multiply these average values by a random number sampled from a normal distribution with a mean 0.4 and a standard deviation 0.2. Unit manufacturing and engineering costs ($c^f_j$ and $c^e_j$) are generated from discrete uniform distributions between [1, 10] and [50, 100] respectively. Finally, the total available budget is calculated by multiplying the cost of using all factory and engineering resources in all periods by $Bc$, which we refer to as the budget fraction coefficient, i.e. $Bc\times \sum_{j \in \mathcal{J}}\sum_{t\in \mathcal{T}}(R^f_tc^f_j+R^e_tc^e_j)$. 


\subsection{Performance of the Algorithm}
We analyze the performance of the proposed CCG algorithm on 36 instance classes presented in Table~\ref{tab:par_level} with the budget coefficient $Bc=0.4$. 
 We generate five instances of each class, resulting in a total of $36\times 5= 180$ instances. We set the time limit to 3 hours, and the time limit for solving the master problem to 2 hours in each iteration of the CCG algorithm.  We allow the master problem to be solved in the given time limit even if the total solution time limit (i.e. 3 hours) is exceeded. The algorithm is implemented in the Python programming language with the Gurobi 10.0.0 solver. All instances are run on 
the MIT Supercloud system with an Intel Xeon Platinum 8260 processor with 48 cores
and 192 GB of RAM \citep{MITSupercloud2018}. 

\begin{table} 
    \centering
    \caption{Set of values and distributions considered to generate instances}
    \label{tab:par_level}
    \begin{tabular}{l|l}
    \toprule
    Parameter or coefficient & set of values or distributions\\
    \midrule
    $T$ &  $\{8,12,16\}$\\
    $J$& $\{2,4,8\}$\\
    $N_j \qquad \forall j\in \mathcal{J}$&\{8,12\}\\
    $P_j\qquad \forall j \in \mathcal{J}$& $\{2,6\}$\\
    \bottomrule
    \end{tabular}
\end{table}



\begin{figure}[htbp]
    \centering
    \includegraphics[width=\textwidth]{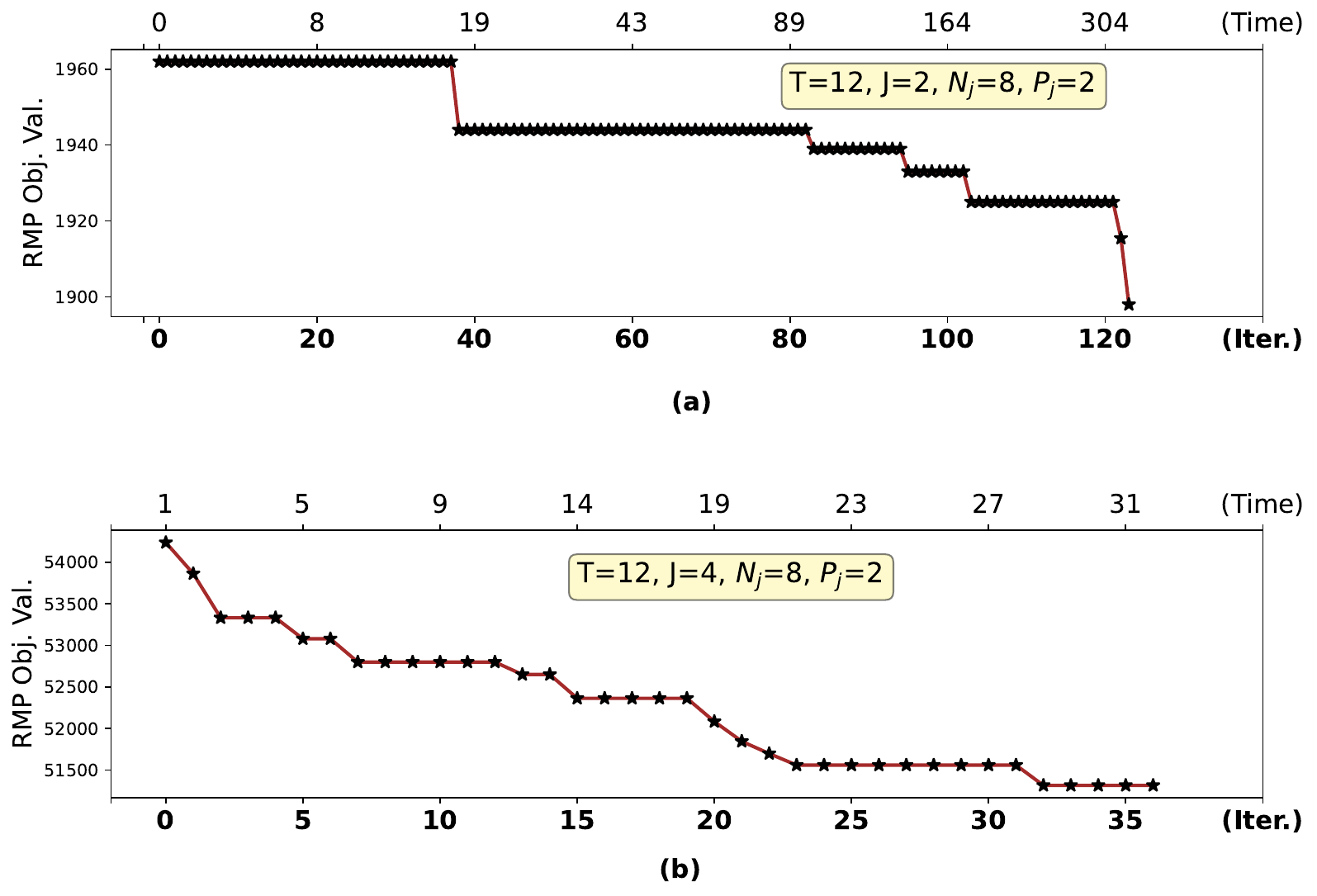}
    \caption{Convergence of the algorithm in some of the instances}
    \label{fig:alg-conv}
\end{figure}

Table~\ref{tab:alg-perform} reports the performance of the solution approach on these randomly generated instances. The solution time and number of iterations are only reported for instances that are solved within the time limit. 
Instance class C22 has the highest average CPU time of more than 2 hours, and C5  requires on average 50 iterations to achieve the optimal solution, the highest number of iterations among all classes. The algorithm fails to solve instances in classes C24 and  C36. These instances have 8 PDs each with 12 total products, and hence represent relatively large instances. In total, 48 out of 180 instances remain unsolved within the 3-hour time limit. 

The size of the instances grows with the number of periods, PDs, total products, and new products. Across all instances, the average solution time increases from 71 seconds for instances with $T=8$ to 970 seconds for instances with $T=12$. However, the average solution time slightly decreases to 815 seconds when $T=16$.  This is because, in practice, large $T$ requires coordination over a longer period of time which could potentially increase the problem complexity. However, larger $T$ also provides PDs with more scheduling flexibility to develop products. Therefore, the impact of this parameter on algorithm performance may not be monotone. The number of unsolved instances remains the same across different values of the planning horizon length $T$. The number of PDs shows a similar trend to the number of periods in terms of solution time. However, the number of unsolved instances grows from 0 when $J=2$ to 13 and 35 when $J=4$ and $J=8$, respectively. Therefore, the number of PDs has more pronounced impact on the problem complexity. \bt{This is also intuitive, since a larger number of PDs increases the likelihood that several PDs are seeking to introduce new products in the same time interval, increasing contention for manufacturing and engineering resources.} The number of total products contributes only slightly to the solution time of instances, whereas the number of new products greatly increases solution time from 47 to 1307 seconds.  This is expected as adding more followers not only increases model size but also requires coordination among more entities. Overall, the number of PDs and new products have the most impact on the computational burden.

\subsection{Convergence of the Algorithm}
The convergence of the proposed CCG algorithm for 2 instances whose optimal solution is reached after more than 35 iterations is depicted in Figure~\ref{fig:alg-conv}, which shows the value of the RMP objective throughout the iterations. The algorithm does not show a sign of ``tailing-off'', i.e.,  the convergence is relatively steady in plot (b) or even faster at the end of plot (a). However, in both instances, the algorithm is unable to improve the RMP objective for several iterations. This is because different budget allocations can still result in the same engineering and manufacturing resource allocations. This observation suggests that an algorithm that can use the generated cut and columns to remove solutions with the same objective value may potentially accelerate the convergence. 

\subsection{Impact of Budget}
\begin{figure}[htbp]
    \centering
    \includegraphics[width=\textwidth]{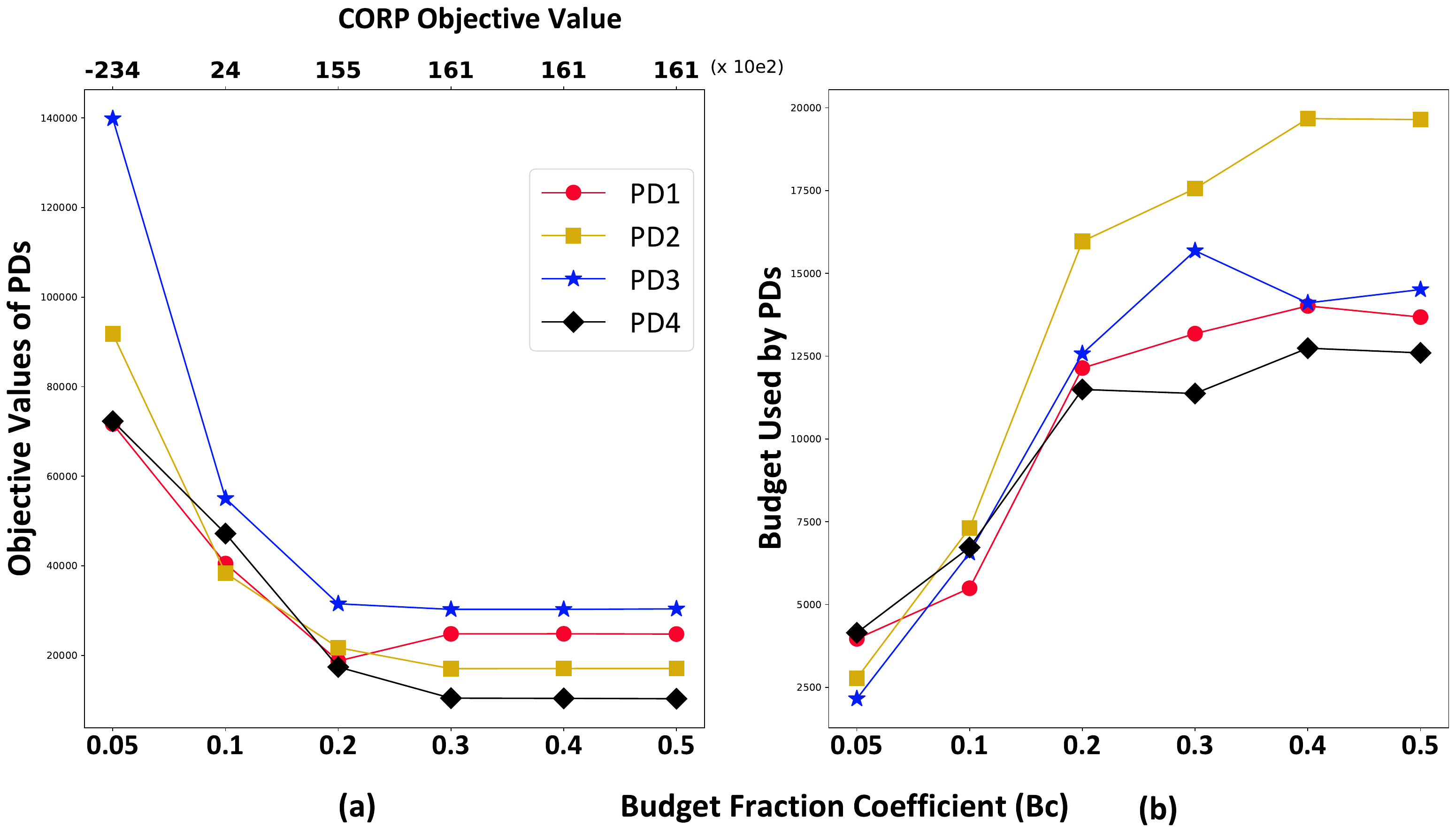}
    \caption{Impact of Budget on the objective function value of the CORP and PDs. The right plot shows PDs' objective value in y-axis and the CORP's objective value corresponding to each $Bc$ in the top x-axis. The left plot shows the budget used by PDs on the y-axis.}
    \label{fig:budget-impact}
\end{figure}

Since CORP uses budget allocation to influence the decisions of the followers, this section examines the impact of the budget on the optimal objective function of CORP  and the PDs, and the performance of the algorithm. We consider instance C41 (i.e, $T=12, J=4, N_j=10, P_j=4$) and let budget fraction coefficient $Bc$ take values from the set $\{0.05, 0.1, 0.2, 0.3, 0.4, 0.5\}$.
Figure~\ref{fig:budget-impact} illustrates the PDs' and CORP's objective function values (left plot) and the amount of budget consumed by each PD (right plot). The top axis in Figure~\ref{fig:budget-impact}a shows the objective value of CORP over different $Bc$ levels.  Increasing the budget available to CORP initially improves (reduces) PDs' costs while increasing CORP's objective.
However, the objective function of each decision unit remains unchanged for $Bc \ geq 0.3$, implying that all product development and manufacturing activities have enough engineering and manufacturing resources. \bt{Different PDs receive different budget allocations, which is intuitive: different PDs can be expected to contribute in differing amounts to the firm's profit based on the markets they serve and the products they produce, so the PDs generating the highest return on the investment represented by their budget allocation ought to receive more budget. An interesting observation, however, is that for values of $Bc \ geq 0.3$, the PD's objective function values remain unchanged but the amount of budget each PD consume still varies with the value of $Bc$. Given that the objective function values of each PD remain constant over this range of $Bc$ values, this may suggest that there are alternative optimal solutions for the PDs that yield the same objective value but consume budget in different ways.}
Our model can be used to analyze such limits to find the optimal funding level of PDs.



\subsection{Impact of Competition}
\begin{figure}
    \centering
    \includegraphics[width=0.9\textwidth]{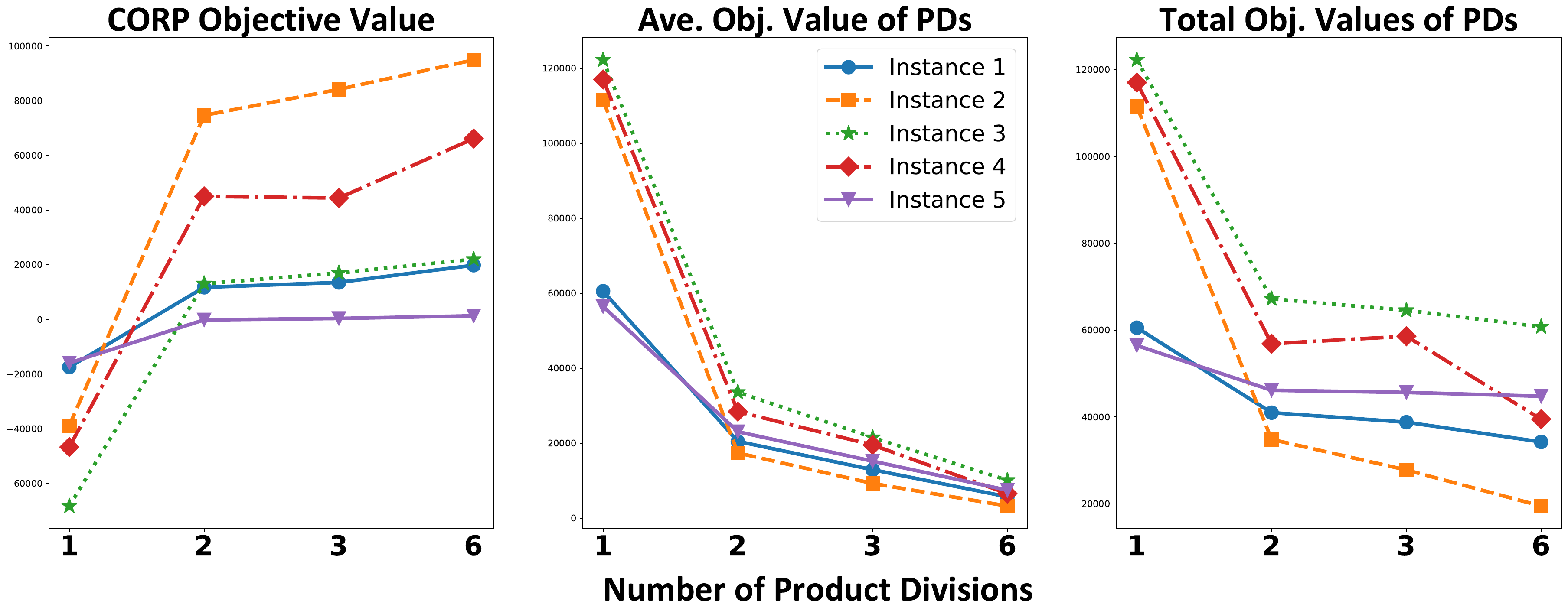}
    \caption{Impact of competition}
    \label{fig:comp-impact}
\end{figure}

There is more competition for factory and engineering resources as the number of PDs increases. We generate four instance classes with 12 total and 6 new products in Table~\ref{tab:competition}. The parameters of these products are the same across instances, hence the differences in the objective functions are due to changes in the number of PDs. We run 5 instances of each class and show CORP's objective value and the average objective value of PDs in Figure~\ref{fig:comp-impact}. For all instances, CORP's objective function value improves as the number of PDs grows while the average objective values of individual followers as well as the sum of the objective values of PDs decreases. The only exception is instance 4 when the number of PDs increases from 2 to 3. Further analyses of cost components show that the backorder cost dominates production and inventory costs, thus driving the behavior of planning entities under competition.

\begin{table}[H]
    \centering
    \begin{tabular}{l|l l l l }
    \toprule
         Instance Class& $T$ & $J$ & $N_j$ & $P_j$ \\
         \midrule
         1& 12 & 6 & 2 & 1 \\
         2&12&3&4&2\\
         3&12&2&6&3\\
         4&12&1&12&6\\
         \midrule 
    \end{tabular}
    \caption{Instance classes to evaluate the impact of competition}
    \label{tab:competition}
\end{table}


\section{Extensions of the Model}
\bt{An interesting extension to the model is to allow the explicit prioritization by corporate management of ccertain PDs over others. A common situation in which this occurs is when a firm identifies a particular market as critical to its operations in some way, to the extent that they are willing to support the PD operating in that market despite its being less profitable than other PDs, at least in the short term. One way to approach this would be to solve the problem considering only the prioritized PDs, fixing the resulting decisions, after which the remaining budget can be allocated among remaining PDs by solving a second instance of the problem. By its nature this approach is suboptimal in the short term, and would not be necessary if the planning horizon $T$ were sufficiently long to capture the potential increases in future revenue from the currently unprofitable PDs.}

\bt{A second extension is the incorporation of endogenous deadlines for product development operations, which would allow the explicit modelling of a Product Development unit shared by all PDs as a distinct agent in its own right. This problem is substantially more difficult that the problem addressed here, which represents the lower level as a set of PDs with identically structured constraints and objective functions that vary only in their parameter values. In particular, the PDs are now constrained by the decisions of both CORP and the product engineering unit, destroying the two-level hierarchical structure upon which our solution algorithms are based. }


\section{Conclusion} \label{sec:Conclusion}
This paper proposes a mixed-integer bilevel model with interdependent followers for the decentralized and hierarchical decision process of managing product transitions in high-technology industries. The model coordinates the allocation of budget and distribution of factory and engineering resources.  The leader of the model is Corporate Management, while followers are Product Divisions (PDs) who are responsible for manufacturing and developing new products. The leader allocates the budget among followers to maximize total system's profit. We present a single-level reformulation of the problem and develop a solution algorithm based on cut-and-column generation. The proposed framework allows for the flexible distribution of decision authority and the incorporation of additional domain-specific constraints.

The performance of the algorithm is examined by extensive computational analysis over randomly generated instances that differ in key problem parameters such as the number of planning periods, PDs and new products. We also present the impact of various problem parameters on the objective function value of the leader. Moreover, we provide managerial insights by examining the effect of competition on the objective function of each PD and find that the higher number of PDs can increase corporate profit as it can streamline the allocation of budget and incentivize the PDs to find more cost-saving development plans.

\bt{The proposed model provides a general framework to evaluate other realistic assumptions such as market priority in which a subset of PDs and their products have higher priority over others. In this situation, one can solve the model for priority PDs, update the resources and solve the model for the remaining PDs. Some other assumptions can also be added with mild modifications to the model. For example, the exogenous deadline for product development is an assumption considered in \citep{KhorramfarEtal2022} and can be incorporated into our model by defining product deadline constraints in the PD models.}

This work can be extended in several directions. Firstly, to deal with larger instances, the algorithm's performance can be enhanced by various speed-up strategies such as warm starting.
Second, this study is conducted under complete information about the value of parameters which is not the case in many practical applications. Therefore, considering uncertainty and information asymmetry in a way that scales up with the tractability of the model would be of great practical interest. Finally, this work can be extended by exploring its relationship with other resource coordination frameworks, such as iterative combinatorial auctions~\citep{BansalEtal2020} and design of mechanisms~\citep{MallikHarker2004}.


\appendix
\section{Additional Results}


\begin{table}[ht]
    \centering
    \caption{The performance of the CCG algorithm}
\begin{tabular}{ c| cccc|rrr|rrr|c}
\toprule
      & && & & \multicolumn{3}{c|}{CPU Time}&\multicolumn{3}{c|}{$\#$ Iterations}&\\
     \cmidrule{6-11}
   Instance Class & $T$ & $J$ & $N_j$ & $P_j$ & mean & min & max  & mean & min & max& $\#$ Unsolved \\
    \midrule 


 1  & 8  & 2 & 8  & 2 & 1    & 0    & 2     & 3   & 1   & 9   & 0 \\
2  & 8  & 2 & 8  & 6 & 17   & 3    & 52    & 3   & 1   & 10  & 0 \\
3  & 8  & 2 & 12 & 2 & 7    & 5    & 8     & 3   & 1   & 7   & 0 \\
4  & 8  & 2 & 12 & 6 & 12   & 10   & 14    & 2   & 1   & 2   & 0 \\
5 & 8  & 4 & 8  & 2 & 148  & 0    & 587   & 51  & 1   & 202 & 1 \\
6 & 8  & 4 & 8  & 6 & 173  & 9    & 416   & 5   & 1   & 19  & 0 \\
7 & 8  & 4 & 12 & 2 & 8    & 6    & 9     & 3   & 1   & 6   & 0 \\
8 & 8  & 4 & 12 & 6 & 158  & 158  & 158   & 9   & 4   & 14  & 3 \\
9 & 8  & 8 & 8  & 2 & 29   & 2    & 53    & 13  & 3   & 32  & 2 \\
10 & 8  & 8 & 8  & 6 & 50   & 50   & 50    & 1   & 1   & 1   & 4 \\
11 & 8  & 8 & 12 & 2 & 11   & 1    & 20    & 7   & 1   & 14  & 3 \\
12 & 8  & 8 & 12 & 6 & 509  & 57   & 961   & 2   & 1   & 3   & 3 \\
13 & 12 & 2 & 8  & 2 & 67   & 1    & 329   & 27  & 1   & 124 & 0 \\
14 & 12 & 2 & 8  & 6 & 13   & 8    & 22    & 2   & 1   & 2   & 0 \\
15 & 12 & 2 & 12 & 2 & 1    & 0    & 1     & 2   & 1   & 5   & 0 \\
16 & 12 & 2 & 12 & 6 & 472  & 11   & 1987  & 6   & 2   & 16  & 0 \\
17 & 12 & 4 & 8  & 2 & 8    & 0    & 32    & 10  & 1   & 37  & 0 \\
18 & 12 & 4 & 8  & 6 & 1902 & 162  & 4741  & 2   & 1   & 5   & 0 \\
19 & 12 & 4 & 12 & 2 & 93   & 1    & 263   & 34  & 1   & 126 & 1 \\
20 & 12 & 4 & 12 & 6 & 4279 & 2142 & 5956  & 3   & 2   & 4   & 2 \\
21 & 12 & 8 & 8  & 2 & 446  & 1    & 1330  & 1   & 1   & 2   & 2 \\
22 & 12 & 8 & 8  & 6 & 7855 & 5618 & 10092 & 5   & 5   & 5   & 3 \\
23 & 12 & 8 & 12 & 2 & 104  & 96   & 111   & 34  & 26  & 42  & 3 \\
24 & 12 & 8 & 12 & 6 & -  & -  & -   & - & - & - & 5 \\
25 & 16 & 2 & 8  & 2 & 3    & 0    & 8     & 3   & 1   & 7   & 0 \\
26 & 16 & 2 & 8  & 6 & 88   & 3    & 290   & 2   & 1   & 3   & 0 \\
27 & 16 & 2 & 12 & 2 & 1    & 0    & 2     & 2   & 1   & 4   & 0 \\
28 & 16 & 2 & 12 & 6 & 46   & 17   & 78    & 3   & 2   & 3   & 0 \\
29 & 16 & 4 & 8  & 2 & 6    & 1    & 13    & 8   & 2   & 16  & 2 \\
30 & 16 & 4 & 8  & 6 & 2370 & 106  & 7424  & 1   & 1   & 2   & 1 \\
31 & 16 & 4 & 12 & 2 & 3    & 1    & 5     & 2   & 1   & 4   & 1 \\
32 & 16 & 4 & 12 & 6 & 4986 & 552  & 7203  & 29  & 1   & 85  & 2 \\
33 & 16 & 8 & 8  & 2 & 44   & 1    & 167   & 12  & 1   & 43  & 1 \\
34 & 16 & 8 & 8  & 6 & 3471 & 77   & 7203  & 1   & 1   & 1   & 2 \\
35 & 16 & 8 & 12 & 2 & 54   & 28   & 81    & 12  & 2   & 19  & 2 \\
36 & 16 & 8 & 12 & 6 & -  & -  & -   & - & - & - & 5\\
\bottomrule
\end{tabular}
\label{tab:alg-perform}
\end{table}

\bibliographystyle{apalike}
\bibliography{Bibliography.bib}

\begin{thebibliography}{}

\bibitem[Altman and Wynter, 2004]{AltmanWynter2004}
Altman, E. and Wynter, L. (2004).
\newblock Equilibrium, games, and pricing in transportation and telecommunication networks.
\newblock {\em Networks and Spatial Economics}, 4(1):7--21.

\bibitem[Arslan et~al., 2018]{ArslanEtal2018}
Arslan, O., Jabali, O., and Laporte, G. (2018).
\newblock Exact solution of the evasive flow capturing problem.
\newblock {\em Operations Research}, 66(6):1625--1640.

\bibitem[Aussel and Sagratella, 2017]{AusselSagratella2017}
Aussel, D. and Sagratella, S. (2017).
\newblock Sufficient conditions to compute any solution of a quasivariational inequality via a variational inequality.
\newblock {\em Mathematical Methods of Operations Research}, 85(1):3--18.

\bibitem[Bansal et~al., 2022]{BansalEtal2022}
Bansal, A., {\"O}zalt{\i}n, O.~Y., Uzsoy, R., and Kempf, K.~G. (2022).
\newblock Coordination of manufacturing and engineering activities during product transitions.
\newblock {\em Naval Research Logistics (NRL)}.

\bibitem[Bansal et~al., 2020]{BansalEtal2020}
Bansal, A., Uzsoy, R., and Kempf, K. (2020).
\newblock Iterative combinatorial auctions for managing product transitions in semiconductor manufacturing.
\newblock {\em IISE Transactions}, 52(4):413--431.

\bibitem[Bard, 2013]{book:Bard1998}
Bard, J.~F. (2013).
\newblock {\em Practical bilevel optimization: algorithms and applications}, volume~30.
\newblock Springer Science \& Business Media.

\bibitem[Bard and Moore, 1990]{BardMoore1990}
Bard, J.~F. and Moore, J.~T. (1990).
\newblock A branch and bound algorithm for the bilevel programming problem.
\newblock {\em SIAM Journal on Scientific and Statistical Computing}, 11(2):281--292.

\bibitem[Basilico et~al., 2020]{BasilicoEtal2020}
Basilico, N., Coniglio, S., Gatti, N., and Marchesi, A. (2020).
\newblock Bilevel programming methods for computing single-leader-multi-follower equilibria in normal-form and polymatrix games.
\newblock {\em EURO Journal on Computational Optimization}, 8(1):3--31.

\bibitem[Beresnev and Melnikov, 2018]{BeresnevMelnikov2018}
Beresnev, V. and Melnikov, A. (2018).
\newblock Exact method for the capacitated competitive facility location problem.
\newblock {\em Computers \& Operations Research}, 95:73--82.

\bibitem[Bolusani and Ralphs, 2022]{BolusaniRalph2022}
Bolusani, S. and Ralphs, T.~K. (2022).
\newblock A framework for generalized {Benders’} decomposition and its application to multilevel optimization.
\newblock {\em Mathematical Programming}, 196(1-2):389--426.

\bibitem[Calvete et~al., 2019]{CalveteEtal2019}
Calvete, H.~I., Dom{\'\i}nguez, C., Gal{\'e}, C., Labb{\'e}, M., and Marin, A. (2019).
\newblock The rank pricing problem: models and branch-and-cut algorithms.
\newblock {\em Computers \& operations research}, 105:12--31.

\bibitem[Cui and Wu, 2017]{CuiWu2017}
Cui, A.~S. and Wu, F. (2017).
\newblock The impact of customer involvement on new product development: Contingent and substitutive effects.
\newblock {\em Journal of Product Innovation Management}, 34(1):60--80.

\bibitem[Dempe, 2018]{Dempe2018}
Dempe, S. (2018).
\newblock {\em Bilevel optimization: theory, algorithms and applications}, volume~3.
\newblock TU Bergakademie Freiberg, Fakult{\"a}t f{\"u}r Mathematik und Informatik.

\bibitem[DeNegre and Ralphs, 2009]{DenegreRalph2009}
DeNegre, S.~T. and Ralphs, T.~K. (2009).
\newblock A branch-and-cut algorithm for integer bilevel linear programs.
\newblock In {\em Operations research and cyber-infrastructure}, pages 65--78. Springer.

\bibitem[Druehl et~al., 2009]{DruehlEtal2009}
Druehl, C.~T., Schmidt, G.~M., and Souza, G.~C. (2009).
\newblock The optimal pace of product updates.
\newblock {\em European Journal of Operational Research}, 192(2):621--633.

\bibitem[Erhun et~al., 2007]{ErhunConcalves2007}
Erhun, F., Concalves, P., and Hopman, J. (2007).
\newblock The art of managing new product transitions.
\newblock {\em Sloan Management Review}, 48(3):73--80.

\bibitem[Fabiani and Grammatico, 2019]{FabianiGrammatico2019}
Fabiani, F. and Grammatico, S. (2019).
\newblock Multi-vehicle automated driving as a generalized mixed-integer potential game.
\newblock {\em IEEE Transactions on Intelligent Transportation Systems}, 21(3):1064--1073.

\bibitem[Facchinei and Kanzow, 2010]{FacchineiKanzow2010}
Facchinei, F. and Kanzow, C. (2010).
\newblock Generalized nash equilibrium problems.
\newblock {\em Annals of Operations Research}, 175(1):177--211.

\bibitem[Ferrer and Swaminathan, 2006]{FerrerSwaminathan2006}
Ferrer, G. and Swaminathan, J.~M. (2006).
\newblock Managing new and remanufactured products.
\newblock {\em Management science}, 52(1):15--26.

\bibitem[Fischetti et~al., 2018]{FischettiEtal2018}
Fischetti, M., Ljubi{\'c}, I., Monaci, M., and Sinnl, M. (2018).
\newblock On the use of intersection cuts for bilevel optimization.
\newblock {\em Mathematical Programming}, 172(1-2):77--103.

\bibitem[Gokpinar et~al., 2010]{GokpinarEtal2010}
Gokpinar, B., Hopp, W.~J., and Iravani, S.~M. (2010).
\newblock The impact of misalignment of organizational structure and product architecture on quality in complex product development.
\newblock {\em Management science}, 56(3):468--484.

\bibitem[Hemmati and Smith, 2016]{HemmatiSmith2016}
Hemmati, M. and Smith, J.~C. (2016).
\newblock A mixed-integer bilevel programming approach for a competitive prioritized set covering problem.
\newblock {\em Discrete Optimization}, 20:105--134.

\bibitem[Karabuk and Wu, 2003]{KarabukWu2003}
Karabuk, S. and Wu, S.~D. (2003).
\newblock Coordinating strategic capacity planning in the semiconductor industry.
\newblock {\em Operations Research}, 51(6):839--849.

\bibitem[Karabuk and Wu, 2005]{KarabukWu2005}
Karabuk, S. and Wu, S.~D. (2005).
\newblock Incentive schemes for semiconductor capacity allocation: A game theoretic analysis.
\newblock {\em Production and Operations Management}, 14(2):175--188.

\bibitem[Khorramfar et~al., 2022]{KhorramfarEtal2022}
Khorramfar, R., {\"O}zalt{\i}n, O.~Y., Kempf, K.~G., and Uzsoy, R. (2022).
\newblock Managing product transitions: A bilevel programming approach.
\newblock {\em INFORMS Journal on Computing}, 34(5):2828--2844.

\bibitem[Kleinert et~al., 2021]{KleinertEtal2021_survey}
Kleinert, T., Labb{\'e}, M., Ljubi{\'c}, I., and Schmidt, M. (2021).
\newblock A survey on mixed-integer programming techniques in bilevel optimization.
\newblock {\em EURO Journal on Computational Optimization}, 9:100007.

\bibitem[Koca et~al., 2010]{KocaEtal2010}
Koca, E., Souza, G.~C., and Druehl, C.~T. (2010).
\newblock Managing product rollovers.
\newblock {\em Decision Sciences}, 41(2):403--423.

\bibitem[Lavigne et~al., 2000]{LavigneSavard2000}
Lavigne, D., Loulou, R., and Savard, G. (2000).
\newblock Pure competition, regulated and stackelberg equilibria: application to the energy system of {Q}uebec.
\newblock {\em European Journal of Operational Research}, 125(1):1--17.

\bibitem[Le{\'o}n and Farris, 2011]{LeonEtal2011}
Le{\'o}n, H. C.~M. and Farris, J.~A. (2011).
\newblock Lean product development research: Current state and future directions.
\newblock {\em Engineering Management Journal}, 23(1):29--51.

\bibitem[Li et~al., 2014]{LiEtal2014}
Li, H., Graves, S.~C., and Huh, W.~T. (2014).
\newblock Optimal capacity conversion for product transitions under high service requirements.
\newblock {\em Manufacturing \& Service Operations Management}, 16(1):46--60.

\bibitem[Li et~al., 2010]{LiEtal2010}
Li, H., Graves, S.~C., and Rosenfield, D.~B. (2010).
\newblock Optimal planning quantities for product transition.
\newblock {\em Production and Operations Management}, 19(2):142--155.

\bibitem[Liang et~al., 2014]{LiangEtal2014}
Liang, C., {\c{C}}akany{\i}ld{\i}r{\i}m, M., and Sethi, S.~P. (2014).
\newblock Analysis of product rollover strategies in the presence of strategic customers.
\newblock {\em Management Science}, 60(4):1033--1056.

\bibitem[Lobel et~al., 2015]{LobelEtal2015optimizing}
Lobel, I., Patel, J., Vulcano, G., and Zhang, J. (2015).
\newblock Optimizing product launches in the presence of strategic consumers.
\newblock {\em Management Science}, 62(6):1778--1799.

\bibitem[Lozano and Smith, 2017]{LozanoSmith2017}
Lozano, L. and Smith, J.~C. (2017).
\newblock A value-function-based exact approach for the bilevel mixed-integer programming problem.
\newblock {\em Operations Research}, 65(3):768--786.

\bibitem[Mallik and Harker, 2004]{MallikHarker2004}
Mallik, S. and Harker, P.~T. (2004).
\newblock Coordinating supply chains with competition: Capacity allocation in semiconductor manufacturing.
\newblock {\em European Journal of Operational Research}, 159(2):330--347.

\bibitem[{\"O}zalt{\i}n et~al., 2018]{OzaltinEtal2018}
{\"O}zalt{\i}n, O.~Y., Prokopyev, O.~A., and Schaefer, A.~J. (2018).
\newblock Optimal design of the seasonal influenza vaccine with manufacturing autonomy.
\newblock {\em INFORMS Journal on Computing}, 30(2):371--387.

\bibitem[Rash and Kempf, 2012]{RashKempf2012}
Rash, E. and Kempf, K. (2012).
\newblock Product line design and scheduling at {Intel}.
\newblock {\em Interfaces}, 42(5):425--436.

\bibitem[Reuther et~al., 2018]{MITSupercloud2018}
Reuther, A., Kepner, J., Byun, C., Samsi, S., Arcand, W., Bestor, D., Bergeron, B., Gadepally, V., Houle, M., Hubbell, M., et~al. (2018).
\newblock Interactive supercomputing on 40,000 cores for machine learning and data analysis.
\newblock In {\em 2018 IEEE High Performance extreme Computing Conference (HPEC)}, pages 1--6. IEEE.

\bibitem[Sagan, 2020]{book:Sagan2020}
Sagan, B.~E. (2020).
\newblock {\em Combinatorics: The art of counting}, volume 210.
\newblock American Mathematical Soc.

\bibitem[Sagratella, 2017]{Sagratella2017a}
Sagratella, S. (2017).
\newblock Algorithms for generalized potential games with mixed-integer variables.
\newblock {\em Computational Optimization and Applications}, 68(3):689--717.

\bibitem[Sagratella et~al., 2020]{SagratellaSchmidt2020}
Sagratella, S., Schmidt, M., and Sudermann-Merx, N. (2020).
\newblock The noncooperative fixed charge transportation problem.
\newblock {\em European Journal of Operational Research}, 284(1):373--382.

\bibitem[Shi et~al., 2007]{ShiEtal2007}
Shi, C., Zhou, H., Lu, J., Zhang, G., and Zhang, Z. (2007).
\newblock The kth-best approach for linear bilevel multifollower programming with partial shared variables among followers.
\newblock {\em Applied Mathematics and Computation}, 188(2):1686--1698.

\bibitem[Sundaramoorthy et~al., 2012]{SundaramoorthyEtal2012}
Sundaramoorthy, A., Evans, J.~M., and Barton, P.~I. (2012).
\newblock Capacity planning under clinical trials uncertainty in continuous pharmaceutical manufacturing, 1: mathematical framework.
\newblock {\em Industrial \& engineering chemistry research}, 51(42):13692--13702.

\bibitem[Tavasl{\i}o{\u{g}}lu et~al., 2019]{tavasliouglu2019solving}
Tavasl{\i}o{\u{g}}lu, O., Prokopyev, O.~A., and Schaefer, A.~J. (2019).
\newblock Solving stochastic and bilevel mixed-integer programs via a generalized value function.
\newblock {\em Operations Research}, 67(6):1659--1677.

\bibitem[Wang and Xu, 2017]{WangXu2017}
Wang, L. and Xu, P. (2017).
\newblock The watermelon algorithm for the bilevel integer linear programming problem.
\newblock {\em SIAM Journal on Optimization}, 27(3):1403--1430.

\bibitem[Wang et~al., 2021]{Wang2Etal021}
Wang, Z., Liu, F., Ma, Z., Chen, Y., Jia, M., Wei, W., and Wu, Q. (2021).
\newblock Distributed generalized nash equilibrium seeking for energy sharing games in prosumers.
\newblock {\em IEEE Transactions on Power Systems}.

\bibitem[Wu et~al., 2009]{WuEtal2009}
Wu, L., De~Matta, R., and Lowe, T.~J. (2009).
\newblock Updating a modular product: How to set time to market and component quality.
\newblock {\em IEEE Transactions on Engineering Management}, 56(2):298--311.

\bibitem[Wu et~al., 2010]{wu2010improving}
Wu, S.~D., Kempf, K.~G., Atan, M.~O., Aytac, B., Shirodkar, S.~A., and Mishra, A. (2010).
\newblock Improving new-product forecasting at {Intel} corporation.
\newblock {\em Interfaces}, 40(5):385--396.

\bibitem[Xu and Wang, 2014]{XuWang2014}
Xu, P. and Wang, L. (2014).
\newblock An exact algorithm for the bilevel mixed integer linear programming problem under three simplifying assumptions.
\newblock {\em Computers \& operations research}, 41:309--318.

\bibitem[Yue et~al., 2019]{YueEtal2019}
Yue, D., Gao, J., Zeng, B., and You, F. (2019).
\newblock A projection-based reformulation and decomposition algorithm for global optimization of a class of mixed integer bilevel linear programs.
\newblock {\em Journal of Global Optimization}, 73(1):27--57.

\bibitem[Zhao et~al., 2023]{ZhaoEtal2023}
Zhao, D., Coyle, S., Sakti, A., and Botterud, A. (2023).
\newblock Market mechanisms for low-carbon electricity investments: A game-theoretical analysis.
\newblock {\em IEEE Transactions on Energy Markets, Policy and Regulation}.

\end{thebibliography}
\end{document}